\documentclass[twocolumn, pre, amsmath, amssymb, letterpaper, superscriptaddress, showpacs]{revtex4}

\usepackage{graphicx}
\usepackage{dcolumn}
\usepackage{bm}
\usepackage{amsmath}
\usepackage{color}

\definecolor{gray}{rgb}{0.5,0.5,0.5}
\definecolor{red2}{rgb}{1.0,0.25,0.25}

\begin{document}


\title{Reconstructing Interaction Potentials in Thin Films from Real-Space Images}

\author{Jonas Gienger}
  \affiliation{%
    Department of Physics, Humboldt-Universit\"{a}t zu Berlin,
    Newtonstra\ss{}e 15, 12489 Berlin, Germany}%
  \affiliation{%
    Physikalisch-Technische Bundesanstalt (PTB),
    Abbestra\ss{}e 2--12, 10587 Berlin, Germany}%
\author{Nikolai Severin}
  \affiliation{%
    Department of Physics, Humboldt-Universit\"{a}t zu Berlin,
    Newtonstra\ss{}e 15, 12489 Berlin, Germany}%
\author{J\"{u}rgen P. Rabe}
  \affiliation{%
    Department of Physics, Humboldt-Universit\"{a}t zu Berlin,
    Newtonstra\ss{}e 15, 12489 Berlin, Germany}%
  \affiliation{%
  IRIS Adlershof, Humboldt-Universit\"{a}t zu Berlin, 
  Zum Gro\ss{}en Windkanal 6, 12489 Berlin, Germany}
\author{Igor M. Sokolov}%
\email{sokolov@physik.hu-berlin.de}
  \affiliation{%
    Department of Physics, Humboldt-Universit\"{a}t zu Berlin,
    Newtonstra\ss{}e 15, 12489 Berlin, Germany}%

\date{\today}

\begin{abstract}
 We demonstrate that an inverse Monte Carlo approach allows to reconstruct effective interaction potentials from real-space images. 
 The method is exemplified on monomolecular ethanol-water films imaged with scanning force microscopy (SFM), which provides the spatial distribution of the molecules.
 Direct Monte Carlo simulations with the reconstructed potential allow for obtaining
 characteristics of the system which are unavailable in the experiment, such as the heat capacity of the monomolecularly thin film, and for
 a prediction of the critical temperature of the demixing transition.
\end{abstract}

\maketitle

Real-space imaging is widespread in physical, chemical and biological sciences and is often the only or the best way to obtain information about a system. 
Examples include, but are not limited to scanning probe or optical microscopies. A particular challenge is to predict the behavior of an experimental system outside of the experimentally accessible set of parameters 
such as time or temperature. This can be achieved given the knowledge of the interactions in the system, which can indeed be restored from real-space images, as
we discuss in the present work.

This work is motivated by recent investigations of a monomolecular film of nanophase separated water-ethanol mixtures confined in a slit-pore between a mica substrate and a graphene coating
as observed by scanning force microscopy (SFM) \cite{severin2014dynamics, severin2015nanophase}.
%
The system exemplifies a situation where the distribution of species is known experimentally, yet the
interactions between them are difficult to estimate. An effective interaction emerges from the complex
interplay of intermolecular forces, graphene's elastic energy, and (screened) dipole-dipole
interactions due to charge transfer \cite{severin2015nanophase}. In Ref.~\cite{severin2015nanophase} it has only been possible to 
provide a qualitative discussion, which 
has shown that the experimental situation 
resembles the behavior of a system with short-range attractive and long-range repulsive interactions 
between  molecules of the same type, at a temperature above the critical point.
However, such a discussion could give neither the strength nor the range of corresponding interactions.
Now we use this system as an example for the reconstruction of interaction potentials from real-space images,
which gives the key for a prediction of properties which were not immediately observed, including the critical
temperature for the phase separation. The ability to estimate the critical temperature is of crucial importance
for designing new experiments on this interesting system. 

\begin{figure}[t]
\pdfpxdimen=\dimexpr 1in/96\relax
  \centering
  \setlength\fboxsep{0pt}
  \setlength\fboxrule{1pt}
  {\color{gray}\fbox{\includegraphics[clip=true, trim = 0px 0px 0px 0px, width=196pt]{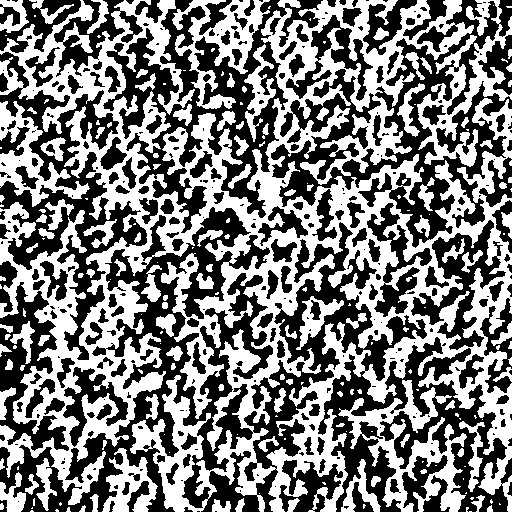}}}
  \\[-54pt]\hspace{141pt}
  \setlength\fboxrule{3pt}
  {\color{red2}\fbox{\includegraphics[clip=true, trim = 0px 0px 0px 0px, width=49pt]{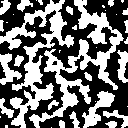}}}
\caption{(Color online)
Binarized SFM topography image:
White (black) pixels are ethanol-rich (water-rich) areas, respectively.
Single-layer graphene, $512	\times512$ pixel, $500\,\mathrm{nm}$ wide. 
See \cite{severin2015nanophase} for more details. The inset (red frame) shows a simulated configuration 
with an effective potential from IMC on a $128 \times 128$ lattice (periodic boundaries).}
\label{fig:exp_image}
\end{figure}

The SFM image of a system is a rasterized and pixelled image representing the height profile of the graphene film.
The height profile observed shows a clear bimodal structure, well-approximated by two Gaussian peaks 
of similar widths \cite{severin2015nanophase}; the finite width can be attributed to thermal and instrumental noise.
The difference in height corresponding to the peaks of the distribution is around 1\,{\AA} and is of the order 
of the difference in size of water and ethanol molecules. The corresponding domains are interpreted as water-rich and alcohol-rich clusters.
The  SFM images are therefore filtered and reduced to 1\,bit color depth
(see \cite{severin2015nanophase} for details). 
One of such images is shown in Fig.~\ref{fig:exp_image}. 
Here the system is in thermodynamic equilibrium.
%

The present work is devoted to the question: Which {\em quantitative} information can be drawn from
images as the one in Fig.~\ref{fig:exp_image}? Our approach is based on the determination of an effective 
pair interaction potential directly from the real-space images by use of the inverse Monte Carlo (IMC) method.

The pixelled SFM images trivially map to square lattices, and hence a lattice binary mixture model suggests itself for a coarse-grained description
of the situation (the coarse graining in experiment is a trivial consequence of the relatively low lateral resolution of the SFM, being of the 
order of a few nanometers, so that single molecules are not resolved).
The model is equivalent to a lattice spin system \cite{Stanley}. 
Such spin models (Ising models with additional long-ranged potentials) have been
discussed in the literature and simulated, see, e.\,g., \cite{PhysRevLett.75.950}. 
We assign the values $\sigma_i=+1$ ($\sigma_i=-1$) to the white (black)
pixels of the image. The translation from the language of binary mixture into the one of lattice spin models is discussed in Appendix~\ref{app:binary_mixture}.
IMC is a well-established method to calculate effective pair interaction 
potentials from pair correlation functions (CFs), see Ref.~\cite{Toth2007overview} for a review,
and is typically applied in a continuous setting.

Moreover, we pose the IMC method into a larger scope of maximum likelihood estimates. This   can allow 
in future for its application to more complex (say, anisotropic) situations, to situations under large experimental errors, and 
for the comparison of different models using information criteria.
For the discussion of the maximal likelihood approaches in statistical physics see, e.\,g., Ref.~\cite{Mastromatteo} and references therein.
This work is however devoted to a different popular inverse problem of statistical physics, the inverse Ising model.

As discussed in \cite{severin2015nanophase}, the effective interaction between the molecules in a slit-pore between the mica surface and the graphene
film contains essentially three contributions: The direct van-der-Walls interaction between the molecules, the effective electrostatic interaction due to the charge transfer, and the additional
interaction due to the elastic deformation of the graphene sheet pushed towards the sublayer by the Casimir force and conforming to the molecular relief. 
The third contributions might contain non-additive (multiparticle) components, 
which are however short-ranged
(the elastic interaction potential behaves like $1/r^4$ \cite{marchenko2002elastic, Peyla2003elastic})
and decay considerably at the distances
of the order of the lateral resolution of the image. Therefore, in our coarse-grained picture, one can, at least as the first approximation, consider only
the pairwise interactions. The Hamiltonian of a spin system for a given spin configuration $\{\sigma\}$
is
\begin{equation}
 \mathcal{H}(\{\sigma\}) =
  \sum_{i,j\in\mathcal{G}}W(\bm{r}_{ij})\, \sigma_i\, \sigma_j- H_\text{ext}\,M(\{\sigma\}), \label{eq:Hamiltonian}
\end{equation}
where 
$W$ is the  effective pair potential, $H_\text{ext}$ is an external magnetic field and 
$M(\{\sigma\}) = \sum_{i\in\mathcal{G}}\sigma_i$ is the total magnetization of the lattice.
Here $\bm{r}_{ij}$ is the distance vector between pixels $i,j$; $\mathcal{G}$ is the set of all pixels/lattice sites.

We moreover introduce the spin-spin correlation function at some distance vector $\bm{r}$ on the lattice
\begin{equation}
 C_{\bm{r}}(\{\sigma\}) = \frac{1}{N}\sum_{i,j\in\mathcal{G}|\bm{r}_{ij}=\bm{r}}\sigma_i\,\sigma_{j},	\label{eq:correlation_def}
\end{equation}
where $N=|\mathcal{G}|$ is the number of pixels/lattice sites.
Now we can rewrite the Hamiltonian as
\begin{equation}
 \mathcal{H}(\{\sigma\}) =
 N \,\sum_{\bm{r}} W_{\bm{r}} \,C_{\bm{r}}(\{\sigma\}) - H_\text{ext}\,M(\{\sigma\}), \label{eq:Hamiltonian_corr}
\end{equation}
with $ W_{\bm{r}} = W(\bm{r})$.  
Eq.~\eqref{eq:Hamiltonian_corr} shows that the energy of the system
depends only on the correlation function $C$ and the magnetization $M$. 
The case $H_\text{ext} = 0$ corresponds to symmetric black-white coverage, which is (approximately) exhibited by many -- but not all -- images.
The possibility to include  $H_\text{ext} \neq 0$ is discussed in 
Appendix~\ref{app:details_implementation}.

The probability of a given spin configuration $\{\sigma\}$ for a system with known interaction potential  $W$
is given by the Gibbs distribution 
\begin{equation}
 p(\{\sigma\}|W) = \frac{ \exp\left[-\beta \mathcal{H}_W(\{\sigma\})\right]}{Z_W},
 \label{eq:Distribution}
\end{equation}
where $\mathcal{H}_W(\{\sigma\}))$ is the energy of $\{\sigma\}$ for given $W$, $\beta=1/k_\mathrm{B}T$, and $Z_W$ is the corresponding partition function defined as the sum 
$\sum_{\{\sigma\}}  \exp\left[-\beta \mathcal{H}_W(\{\sigma\})\right]$ over all spin configurations. 
Substituting Eq.~\eqref{eq:Hamiltonian_corr} into Eq.~\eqref{eq:Distribution} we see that the 
distribution $p(\{\sigma\}|W)$ is of the exponential class, and that (for $H_\text{ext} = 0$) $C_{\bm{r}}$ is a sufficient statistics, i.\,e.,
gives all information necessary to infer $W_{\bm{r}}$, see, e.\,g., \cite{Mandelbrot}.

We now consider $W$ as a vector of parameters (with entries $W_{\bm{r}}$), and introduce the
likelihood $\mathcal{L}(W;\{\sigma\}) = p(\{\sigma\}|W)$. The corresponding log-likelihood $L(W)= \ln \mathcal{L}(W;\{\sigma\})$ is then
\begin{equation}
 L(W) = -\beta \mathcal{H}_{W}(\{\sigma\}) - \ln{Z_W} = \beta[F_W - \mathcal{H}_W(\{\sigma\})], 
\end{equation}
where now $F_W$ is the free energy of the spin system for given $W$ at temperature $T$ which is independent of the configuration of spins.
If we assume that the spin configuration observed in the experiment is \emph{representative}, the energy $\mathcal{H}_W$ is well approximated by the internal
energy $E_W$ of the system, in which case $L(W) = \beta (F_W - E_W) = - S_W/k_\mathrm{B}$, where $S_W$ is the system's entropy:
The log-likelihood function for the interaction potential is proportional to the negative value of the entropy
of the system with a corresponding spin configuration under the potential assumed. Its maximization then assumes that the most likely value of the interaction
potential is such, that the experimental configuration of spins provides the maximal information about it. We note that the equivalence of the maximum likelihood and 
\emph{minimum} entropy in statistics is known \cite{kriz1968}. Here, however, the statistical entropy has the
meaning of the thermodynamical entropy pertinent to a representative configuration. 

\begin{figure}[t]
 \centering
\includegraphics[clip = true, trim = 4.8pt 8.8pt 15pt 31pt,scale=0.96]{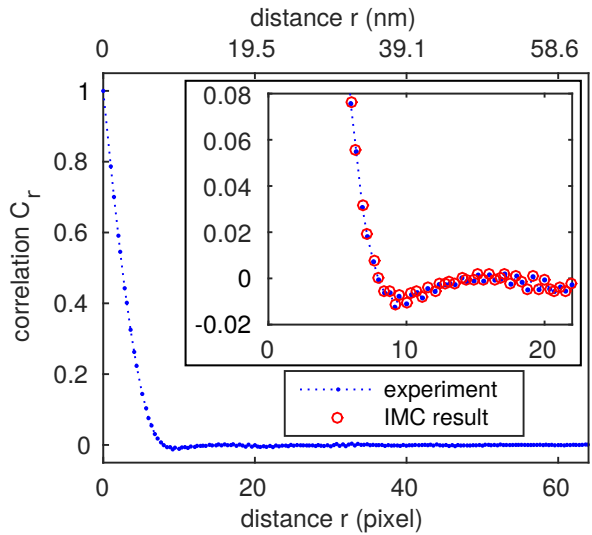}
\caption{(Color online)
 Radial histogram of the ``spin-spin'' CF for the SFM image in Fig.~\ref{fig:exp_image}.
 Inset: A part of the CF used in the calculation and the CF restored by IMC.
}
\label{fig:correlation}
\end{figure}

The representation given by Eq.~\eqref{eq:Hamiltonian_corr} allows for performing an IMC procedure using the 
algorithm proposed in \cite{Lyubartsev1995calculation}, adapted to a lattice system. 
We consider the CF obtained from the experimental image $\hat{C}$, see Fig.~\ref{fig:correlation}, as a vector of data (with entries $\hat{C}_{\bm{r}}$), 
where $\bm{r}$ has been binned radially.
The position of the (global) minimum at $r\approx9\,\rm{px}$ corresponds to a characteristic size of the clusters.
We use $k_\mathrm{B}=1$ in the following.
Let us now fix some specific distance ${\bm{r}}$ and consider the necessary condition for the test potential $W$ 
to maximize $L(W)$ (or to minimize $S_W$) provided the data:
\begin{equation}
\frac{\partial L(W)}{\partial  W_{{\bm{r}}}} = - \beta \frac{\partial \mathcal{H}_W}{\partial W_{\bm{r}}} - \frac{1}{Z_W} \frac{\partial Z_W}{\partial W_{{\bm{r}}}} = 0,
\end{equation}
with
\begin{equation}
  -\frac{1}{Z_W} \frac{\partial Z_W}{\partial W_{{\bm{r}}}}= \frac{\beta}{Z_W} \sum_{\{\sigma\}} \frac{\partial \mathcal{H}_W}{\partial W_{\bm{r}}} \exp\left[-\beta \mathcal{H}_W(\{\sigma\})\right]. 
\end{equation}
Since ${\partial \mathcal{H}_W}/{\partial W_{\bm{r}}}=N\,C_{\bm{r}}(\{\sigma\})$,
the sum in the last expression can be rewritten as the average over the canonical distribution, i.\,e., as the canonical 
correlation function $\mathcal{C}_{\bm{r}}({W}) = \langle C_{\bm{r}}(\{\sigma\}) \rangle|_{W} = \sum_{\{\sigma\}} C_{\bm{r}}(\{\sigma\}) p(\{\sigma\}|W)$
for the given interaction potential. The brackets $\langle ...\rangle$ and $\langle ...\rangle|_W$ denote the average over the canonical distribution,
the latter specifying the dependence on  $W$.
We thus see that the necessary condition for the extremum of the likelihood is the system of equations
\begin{equation}
\mathcal{C}_{\bm{r}}(W) = \hat{C}_{\bm{r}} \text{\;\;for all }\bm{r}.
\label{eq:necessary_condition}
\end{equation}
If the canonical CF is obtained by the Monte Carlo (MC) approach, this inverse problem is solved by IMC.
We note that fitting the CF by IMC methods is a known approach  \cite{Lyubartsev1995calculation}. However,
its validity is typically founded in Henderson's theorem \cite{Henderson1974uniqueness} stating that, 
provided only pairwise interaction, the CF defines the interaction potential uniquely, up to the integration constant. 

The necessary condition for the maximum likelihood can be rewritten as $\nabla_{W} S_{W} =0$.
The algorithmic implementation of the IMC starts from taking  some initial value of the potential vector $W^{(0)}$
(close enough to the fix point $W^*$ delivering the minimum entropy). 
For $W=W^{(0)} +\delta W$, we want to determine $\delta W$ such that $\nabla_{W} S_{W}=\nabla_{\delta W} S_{W}=0$ and expand this
expression to the first order in $\delta W$. The ${\bm{p}}$-th component of  $\nabla_{\delta W} S_W$ is given by
\begin{eqnarray}
&& \hspace{-22pt}
\frac{1}{\beta N} \frac{\partial S_{W^{(0)}+\delta W} }{\partial \delta W_{{\bm{p}}}} 
= \hat{C}_{{\bm{p}}} - \mathcal{C}_{{\bm{p}}}(W^{(0)})\nonumber \\
&& \hspace{-6pt}+ \beta\,N\sum_{{\bm{q}}}\delta W_{{\bm{q}}}\, \left.\mathrm{Cov}\left(C_{\bm{p}}(\{\sigma\}),C_{\bm{q}}(\{\sigma\})\right)\right|_{W^{(0)}}
\label{eq:approx_entropy_min_cond}
\end{eqnarray}
with $\left.\mathrm{Cov}(A,B)\right|_{W} = \left.\langle A\,B \rangle\right|_{W} -\left.\langle A \rangle\right|_{W} \left.\langle B \rangle\right|_{W}$.
The data vector $\hat{C}_{\bm{p}}$ is known,  the vector $\mathcal{C}_{\bm{p}}(W^{(0)})$
and the elements of the covariation matrix are obtained in a direct MC simulation for a given potential $W^{(0)}$. 
Note that the emergence of the covariance matrix of $C_{\bm{p}}$ stresses again the nature of the IMC as a
statistical inference method: This matrix is proportional to the Fisher information matrix of the inference problem \cite{Mastromatteo}.

The above system of linear equations \eqref{eq:approx_entropy_min_cond}  is solved numerically, yielding the (presumably small) 
correction terms $\delta W$ to the potential. The first iteration is then $W^{(1)}=W^{(0)}+\delta W$. These updates are repeated until $W^{(n)}$ converges to $W^*$.
The method is basically a multi-dimensional Newton-Raphson method
to solve $\nabla_{W} S_W=0$, where coefficients are obtained in a statistical simulation. In our MC simulations we use the Glauber dynamics (local spin-flip updates) 
that violates the conservation of particle numbers, in contrast to the (slower) Kawasaki dynamics (local particle-exchange updates),
which however makes no difference when only equilibrium properties are discussed.

If $W^{(0)}$ is already close enough to the solution $W^*$ the algorithm converges, but
if it is too far, instabilities occur. To enforce convergence for a wider range of initial potentials,
Ref.~\cite{Lyubartsev1995calculation} proposed scaling the correction terms, i.\,e., using $W^{(n+1)}=W^{(n)}+\Lambda \delta W^{(n)}$ with $\Lambda\in[0,1]$
at iteration $n+1$. 
Our experience for the SFM images shows that $\Lambda=1$ almost always leads to instabilities, while
suitable values lie in the range $\Lambda\in[0.1, 0.5]$. 
In order to keep the computational complexity 
acceptably low 
we considered $W$ to have a rotational symmetry
(i.\,e. the same value 
for all lattice sites at a given separation; the lattice remains, of course, anisotropic)
and introduced a cutoff radius $R_\text{cut}$ that limits the
range of the pair potential, i.\,e., $W_{\bm{r}}=0$ for all $r>R_\text{cut}$.

\begin{figure}[t]
 \centering
  {\includegraphics[clip=true, trim = 5pt 0.5pt 14pt 15.5pt, scale=0.96]{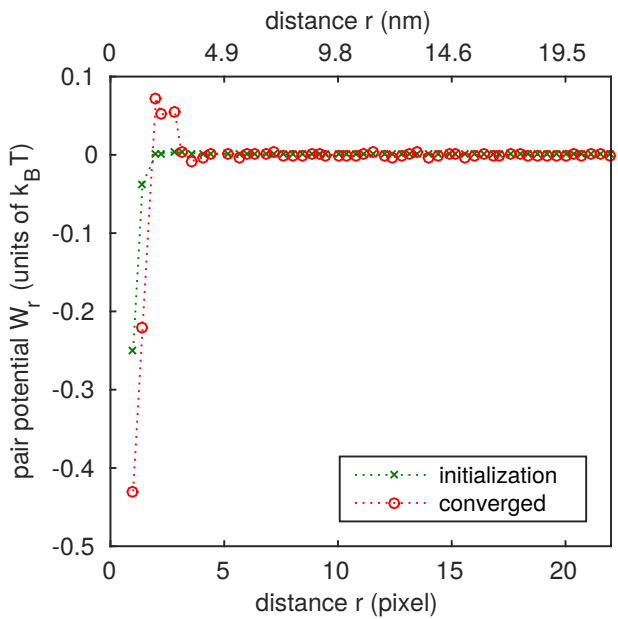}}
  \caption{(Color online)
  Reconstructed potential from the correlation in Fig.~\ref{fig:correlation}.
  Parameters: $64\times64$ lattice, $R_\text{cut}=22$, $H_\text{ext}=-0.001$.
  Hand-tuned initial potential, convergence after 14 inverse MC iterations with progressively increasing measurement times, 
  $7.9\times10^5$ global MC updates in total.
  }
  \label{fig:invMC_example_pot}
\end{figure}

The iterations are stopped when the difference between experimental and simulated correlation becomes small.
To quantify this, we compute the weighted mean square  of $\hat{C} - \mathcal{C}(W)$, 
where the weights are the inverse statistical uncertainties in $\hat{C}$ at the respective distances,
and compare it to unity. All details of implementation are given in Appendix~\ref{app:details_implementation}

For SFM images with a white-pixel concentration of 50\% (or close to it), 
$H_\text{ext}$ can be fixed to 0 (or some suitable small value). For images where this is not the case (finite magnetization), we extended the algorithm 
to further reconstruct the correct value of $H_\text{ext}$.
The extension is in complete analogy to the previous derivation, see Appendix~\ref{app:details_implementation}


We tested the procedure on a two-dimensional ferromagnetic Ising model with additional long-ranged repulsive $1/r^3$ interaction, as discussed before
in the context of ultrathin magnetic films \cite{PhysRevLett.75.950} which exhibit a clustered state at intermediate temperatures. 
The IMC procedure is able to reconstruct the underlying pair-potential from CFs obtained from
a direct simulation, provided sufficient statistical precision of the input data, see also Appendix~\ref{app:test_case}.

The result of the application of the procedure to the experimental CF, Fig.~\ref{fig:correlation}, is shown in Fig.~\ref{fig:invMC_example_pot}.
Inside the cutoff radius $R_\text{cut}$ this potential reproduces the experimental correlation very accurately, cf. the inset in Fig.~\ref{fig:correlation}.
It is attractive at short distances and repulsive at larger ones, in agreement with our qualitative expectations.
We would like to point out that only $\hat{C}_{\bm{r}}$ for $r \leq R_\text{cut}$  (range of inset in Fig.~\ref{fig:correlation})
were used as input data.

\begin{figure}[t]
\centering
  {\includegraphics[clip = true, trim = 1pt 3.5pt 12pt 15pt,scale=0.96]{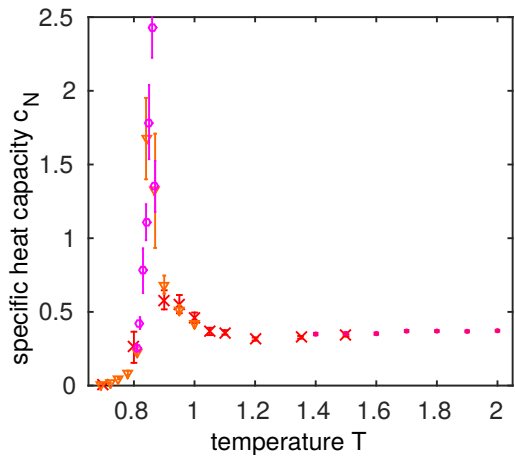}}
 \caption{(Color online)
  MC simulations with reconstructed potential (Fig.~\ref{fig:invMC_example_pot}):  $c_N$ in dependence on  $T$
  with fixed $H_\text{ext}=0$. 
  Measured with decreasing $T$ in four independent series (different markers)
  on a $64\times 64$ lattice, $10^4$ MC updates per measurement
  (of which 10\% for thermalization; $T\in[0.81, 0.87]$ was resolved with $10^5$ MC updates per measurement).  
  }
  \label{fig:reconst_pot_obs_T}
\end{figure}

To restore experimentally relevant properties, like the critical temperature, we used  the reconstructed potential in 
Fig.~\ref{fig:invMC_example_pot} for direct MC simulations on $64\times64$, $128\times128$  and $192\times192$ lattices, giving very similar results.
We examined the specific heat capacity
\begin{equation}
 c_N = \frac{1}{N} \left(\frac{\partial  E }{\partial T}\right)_N
 = \frac{\langle \mathcal{H}^2\rangle-\langle \mathcal{H}\rangle^2}{N\,T^2},
\end{equation}
as well as (specific) magnetization, and magnetic susceptibility. 
The behavior of $c_N (T)$ for $H_\text{ext}=0$ 
is shown in Fig.~\ref{fig:reconst_pot_obs_T} and allows for the definition of the critical temperature.
We chose the temperature scale such that the original experimental temperature is $1\equiv T^*=  298\,\mathrm{K}$, i.\,e., $25\,\mathrm{^\circ C}$.
The phase transition then occurs at $T_c =0.86$, and the results for the magnetization show that the phase at $T<T_c$ is
ferromagnetic, the phase at $T>T_c$ paramagnetic.
Returning to the binary mixture picture, the transition ``paramagnetic--ferromagnetic'' corresponds to ``clustered--demixed''.

To estimate the influence of the cutoff radius $R_{\rm cut}$, we
performed IMC simulations for image G1\#1 using 
the values $R_{\rm cut}=12, 16, \dotsc, 24$ in addition to the case $R_{\rm cut}=22$ discussed above. 
The resulting potentials have the same overall shape, deviations at short distances ($r<10\,{\rm px}$) are small
and hardly visible in a plot. The most significant deviations occur near the respective cutoffs, where we attribute the potentials' fluctuations to statistical errors
anyhow. Thus the exact value of $R_{\rm cut}$ in the IMC algorithm is not crucial.
However, although the potentials differ only slightly, the $T_c$ values obtained from direct MC simulations
scatter somewhat but with no systematic trend. They lie in the range $T_c \in[0.81, 0.86]$, corresponding to a deviation of $6\%\,T_c$.

We further analyzed  images of
double- and triple-layer graphenes, which are shown in Appendix~\ref{app:SFM_images}. The results are qualitatively similar, regarding the potentials themselves (Appendix~\ref{app:IMC_results}) as well as the simulation results: all transitions occured in the
range $T_c \in[0.85, 0.88]$ and no systematic dependence on the number of graphene layers could be observed, see  Appendix~\ref{app:MC_study}, Fig.~\ref{fig:123layers_T_c}.

For the different  images and different $R_{\rm cut}$, $T_c \in[0.81, 0.88]$ was found. Converting this to kelvins, the transitions are 
predicted to occur at $T_c \in [241, 262]\,\mathrm{K}$, i.\,e.,  $[-32, -11]\mathrm{^\circ C}$.
We note that $c_N$ of a monomolecular fluid film cannot be measured on the background of the
large heat capacity of the substrate, but is a very sensitive numerical indicator of the transition, which is in turn observable in the experiment. 
Our estimate for $T_c$ gives a guidance for further experimental work. 

Images of the spin states and plots of correlation functions
show that the clusters grow quickly with temperature decreasing from
$T=1$. At temperatures increasing from $T=1$, the cluster sizes decrease but the system remains in a clustered state at least 
up to $T=2$ and presumably even further. \emph{Nothing special} happens at the experimental temperature $T^*=1$. 

Let us summarize our findings. We used an inverse Monte Carlo approach to reconstruct the effective interactions within a
monomolecularly thin fluid film of a water-ethanol mixture in a slit-pore between  mica and graphene surfaces.
The shape of this
potential qualitatively confirms our qualitative argumentation in \cite{severin2015nanophase}. Direct Monte Carlo simulations
with this potential allow for obtaining information which is not immediately accessible from the experiment, for example, the
critical temperature of the demixing transition.
We considered two-dimensional systems, but the method can be applied to systems of any dimension.

\appendix

\section{Binary Mixture and Lattice Spin Models}\label{app:binary_mixture}

Let us consider the lattice model of a binary mixture, in which each lattice site can be occupied by one
of the species, $A$ or $B$, i.\,e., the variable characterizing the occupation of the site $i$, $q_i$ is either $A$ or $B$.  
The interactions between the different species depends on the distance and therefore 
is given by the set of three distance-dependent interaction energies, $V_{AA}(\bm{r}_{ij})$, $V_{BB}(\bm{r}_{ij})$
and $V_{AB}(\bm{r}_{ij})$, where $\bm{r}_{ij}$ denotes the distance between the corresponding lattice sites.

The energy of the system with given number of $A$ and $B$ sites is then 
\begin{eqnarray}
 \mathcal{H} = &&\sum_{i,j\in\mathcal{G}} \left[ V_{AA}(\bm{r}_{ij}) \delta_{q_i,A}\delta_{q_j,A} + V_{BB}(\bm{r}_{ij}) \delta_{q_i,B}\delta_{q_j,B} \right. \nonumber \\
 && + \left. V_{AB}(\bm{r}_{ij}) (\delta_{q_i,A}\delta_{q_j,B} + \delta_{q_i,B}\delta_{q_j,A}) \right].
\end{eqnarray}
The Kronecker deltas can be replaced by functions of ``spin''-variables $\sigma_i$ equal to $+1$ if the site is occupied by $A$
or $-1$ if it is occupied by $B$. Then $\delta_{q_i,A} = (1+\sigma_i)/2$ and $\delta_{q_i,B} = (1-\sigma_i)/2$.
Substituting Kronecker symbols for these expressions and collecting similar terms we get:
\begin{eqnarray}
 \mathcal{H} &= & \frac{1}{4} \sum_{i,j\in\mathcal{G}} \left[ V_{AA}(\bm{r}_{ij}) +  V_{BB}(\bm{r}_{ij}) + 2 V_{AB}(\bm{r}_{ij}) \right] \nonumber\\
 && + \frac{1}{2} \sum_{i\in\mathcal{G}}  \sigma_i \sum_{j\in\mathcal{G}}  \left[ V_{AA}(\bm{r}_{ij}) -  V_{BB}(\bm{r}_{ij})\right] \\
 && + \frac{1}{4} \sum_{i,j\in\mathcal{G}} \left[ V_{AA}(\bm{r}_{ij}) +  V_{BB}(\bm{r}_{ij}) - 2 V_{AB}(\bm{r}_{ij}) \right] \sigma_i \sigma_j. \nonumber 
\end{eqnarray}
The first term is an offset which does not depend on the configuration and may be neglected (cancels out in the entropy $S$).
Because of the site-independent interaction potential, the sum over $j$ in the second term does not depend on $i$. 
This term, after summing up over all lattice sites $j$ (for large lattices and summable potentials) gives a constant 
multiplied by the sum over $\sigma_i$. The corresponding constant is later incorporated in $H_\text{ext}$: 
\begin{equation}
 H_\text{ext} = \frac{1}{2} \sum_{j\in\mathcal{G}}  \left[ V_{BB}(\bm{r}_{ij}) - V_{AA}(\bm{r}_{ij})  \right] \label{eq:binmixtHext}.
\end{equation}
The third term contains the effective interaction potential, similar in spirit
a Flory-Huggins parameter in the short-range interaction case,
\begin{equation}
 W(\bm{r}_{ij}) = \frac{1}{4} \left[ V_{AA}(\bm{r}_{ij}) +  V_{BB}(\bm{r}_{ij}) - 2 V_{AB}(\bm{r}_{ij}) \right].
\end{equation}
In open systems an additional term, accounting for the energy difference between the molecules in the gas phase and in the pore,
has to be taken into account. This term contributes to the  chemical potential, is linear in the concentrations and is incorporated in
$H_\text{ext}$. Thus, the parameters of the model are a scalar $H_\text{ext}$ and the effective potential $W(\bm{r})$,
for discrete values of $\bm{r}$ corresponding to the distances between the lattice sites.
While we use a lattice spin model in our investigations, the discussion above allow for converting any quantities of interest into 
the corresponding one for a binary mixture.

\section{Details of Implementation}\label{app:details_implementation}
\begin{figure}[t]
 \centering
  \setlength\fboxsep{0pt}
  \setlength\fboxrule{1pt}
  \fbox{\includegraphics[height = 116pt]{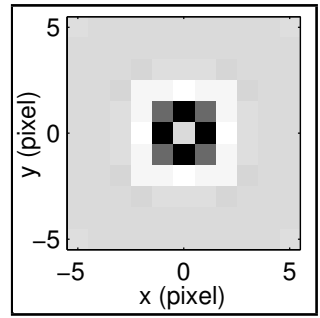}}
  \caption{
  Depiction of the concept of radially symmetric lattice potentials: Lattice plot of the reconstructed potential in Fig.~\ref{fig:invMC_example_pot} (G1\#1 in Fig.~\ref{fig:all_reconstr_pot_SM}).
  Color plot with linear grayscale coding (white = high value).
  }
  \label{fig:invMC_lattice_pot_SM}
\end{figure}

The long-ranged nature of the interaction potentials in the system makes the direct
MC simulation computationally demanding.
In order to keep the complexity of the problem acceptably low already at the stage of direct MC, we introduced a cutoff radius that limits the
(otherwise possibly infinite) range of the pair potential $W$. We denote this cutoff radius by $R_\text{cut}$.
Formally, the potential is $W_r=0$ for distances $r>R_\text{cut}$. Numerically, we do not sum over these distances.
The values of $R_\text{cut}$
used in our simulations are given in Tab.~\ref{tab:invMC_exp}.

In order for equation~\eqref{eq:approx_entropy_min_cond} to be solvable, the covariance-matrix
of the spin-spin correlations must have full rank. Since it is the Hessian matrix of the
entropy $S$ as a function of $W$, this should be the case near a local minimum, where
the matrix is positive definite.
However, if we
let $d$ be the dimension of the system of equations \eqref{eq:approx_entropy_min_cond}, i.\,e., 
the number of independent potential values to be reconstructed,
then the statistical estimates for the covariance matrix (obtained from direct MC) may still be singular if  they are obtained from less than
$d+1$ measurements.
Hence one has to take more than $d$ measurements in each IMC iteration.

As discussed above, the complexity of the algorithm increases with the size of the system of equations \eqref{eq:approx_entropy_min_cond}, i.\,e., with the number of  $W$ values
that are to be reconstructed.
For this reason, one should keep the dimension of the algebraic system low,
in order to have feasible computing times.
Therefore we consider 
{\em radially symmetric potentials}. Of course, exact radial symmetry is not possible on a square lattice. And the radial symmetry
is a {\em postulate} rather than an experimental observation.
Nevertheless we introduce {\em radial binning} of the distances and consider only the radial distance instead of
two-dimensional vectors.
The input data are
correlation functions as shown in Fig.~\ref{fig:correlation} with binned distances. The bins are chosen such that
all nearest neighbors ($r=1$), next-nearest neighbors ($r=\sqrt{2}$), axial next-nearest neighbors ($r=2$), and so on ($r=\sqrt{5},\sqrt{8},3$,\ldots)
fall into an individual bin, respectively. Hence a geometrical meaning is conserved as far as possible. At larger distances bins
are merged if they are close to each other (our criterion: difference $<0.4$).
An example of such a potential is shown in Fig.~\ref{fig:invMC_lattice_pot_SM}.

Since the individual bins have different volumes, the governing equations have to be adapted slightly. With $b_p$ we refer to the volume of bin $p$,
i.\,e., the number of pairs of sites whose distance is inside bin $p$.
The corresponding correlation function is
\begin{equation}
 C_p(\{\sigma\}):=\frac{1}{b_p}\sum_{i,j\in \mathcal{G}|r_{ij} \text{ in bin } p} \sigma_i\,\sigma_j. \label{eq:corr_rad}
\end{equation}
The system of  linear equations \eqref{eq:approx_entropy_min_cond} has to be modified accordingly: $\mathsf{A} \, \bm{x} = \bm{y}$ with
\begin{eqnarray}
 a_{p\,q} &=& \beta\,\,\left.\mathrm{Cov}(b_p\,C_p,b_q\,C_q)\right|_{W},	\label{eq:invMC_SLE_rad_1}
 \\ 
 x_p &=& \delta W_p,	\label{eq:invMC_SLE_rad_2}
 \\
 y_p  &=&  b_p\left[\left.\langle C_p \rangle\right|_{W} - \hat{C}_p\right]	\label{eq:invMC_SLE_rad_3}
\end{eqnarray}
$\forall \; p,q\in\{1,\dotsc,N_B\}$, where $N_B$ is the number of bins.

To set up a suitable convergence criterion, we compute the weighted mean square deviation
of $\langle C_p \rangle|_{W}$ from $\hat{C}_p$, where the
weights are the statistical errors 
of the input data at the corresponding distance $p$. This deviation has to be of the order of unity for
the algorithm to have converged.
But since the  input correlation $\hat{C}$  is computed from a single SFM image
\begin{equation}
 \hat{C}_p = C_p(\{\hat{\sigma}\}),
\end{equation}
no time-series averaging is possible as in a MC simulation, that would allow to obtain a statistical error from fluctuations.

Hence, we need to estimate the statistical error from the single image we have at hand.
We denote the bin averages in equation~\eqref{eq:corr_rad}  by $C_p(\{\hat{\sigma}\})= \langle\langle\hat{\sigma}_i\,\hat{\sigma}_j\rangle\rangle_{r_{ij}\sim p}$.
As an 
estimator for the statistical error of this average, we use  the {\em sample variance}
\begin{eqnarray}
 \left[\Delta \hat{C}_p\right]^2&=& \frac{1}{b_p-1}\left(\langle\langle{\hat{\sigma}_i}^2\,{\hat{\sigma}_j}^2\rangle\rangle_{r_{ij}\sim p}
 - \langle\langle\hat{\sigma}_i\,\hat{\sigma}_j\rangle\rangle_{r_{ij}\sim p}^2\right) \nonumber
  \\&=&  \frac{1}{b_p-1}\left(1-{C_p}(\{\sigma\})^2\right),
\end{eqnarray}
The iterations were stopped when either the prescribed value of the relative error
\begin{equation}
\vartheta = \frac{1}{N_B}\sum_{p=1}^{N_B} \frac{\left(\langle C_p \rangle|_{W} - \hat{C}_p \right)^2}{\left[\Delta \hat{C}_p\right]^2}
, \label{eq:conv_crit}
\end{equation}
was achieved ($\vartheta\le 1$ was used), or after the maximal predefined number of iterations, if the convergence was too slow.
Cases where $\vartheta\le 3$ could not be achieved were considered to not have converged. 
The a-posteriori values of achieved relative errors are given in Tab.~\ref{tab:invMC_exp}.


\subsection*{Extension of the Algorithm}
As already mentioned, the external magnetic field $H_\text{ext}$ can
be included in the algorithm in a fully analogue way as the pair interaction $W$.
The derivative of the Hamiltonian $\mathcal{H}$ with respect to $H_\text{ext}$  is (cf. equation~\eqref{eq:Hamiltonian})
\begin{equation}
 \frac{\partial \mathcal{H}(\{\sigma\})}{\partial H_\text{ext}} = -M(\{\sigma\}).
\end{equation}
To determine also $H_\text{ext}$, one must include the observable $M_S$  as well as its variance and covariance with all
the $C_p$ into the simulation.
The linear system of equations \eqref{eq:invMC_SLE_rad_1}--\eqref{eq:invMC_SLE_rad_3} has to be extended
by
\begin{eqnarray}
  a_{p \,N_B+1} &=& a_{N_B+1 \,p}= \beta\,\,\left.\mathrm{Cov}(b_p\,C_p,M)\right|_{W,H_\text{ext}},\nonumber\\
  a_{N_B+1 \,N_B+1} &=& \beta\,\left.\mathrm{Var}(M)\right|_{W,H_\text{ext}},\\
 x_{N_B+1} &=& -\delta H_\text{ext},\\
 y_{N_B+1}  &=&  \left.\langle M \rangle\right|_{W,H_\text{ext}} - \hat{M}
\end{eqnarray}
$\forall \; p\in\{1,\dotsc,N_B\}$, where $\mathrm{Var}(A) = \mathrm{Cov}(A,A)$ and $\hat{M}$ is the magnetization from the SFM image.
The algorithm will then also provide corrections $\delta H_\text{ext}$ in each iteration, when the system of size $(N_B+1)\times(N_B+1)$ is solved.
Like for the $\delta W$, a damping factor $\Lambda$ has been used with $\delta H_\text{ext}$ and the relative square deviation of the magnetization has been
included in the convergence criterion \eqref{eq:conv_crit}.
One should note, that the corrections $\delta H_\text{ext}$ are obtained not only from the simulated magnetization, but also from the correlations at all distances.

\begin{figure*}[ht!]
 \centering
  \setlength\fboxsep{0pt}
  \setlength\fboxrule{1pt}
  {\includegraphics[clip=true, trim = 4pt 2pt 28pt 11pt, scale=0.96]{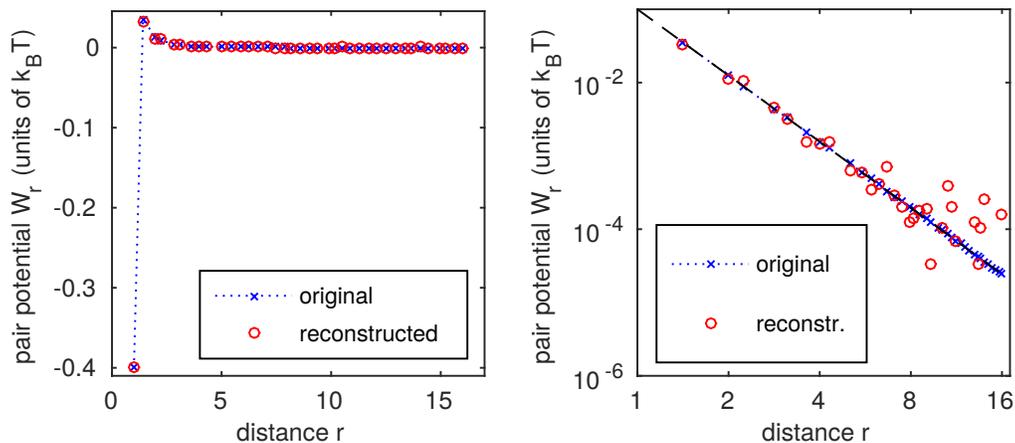}}
 \caption{
 (Color online)
 Reconstructed potential for the ``Ising + dipole-dipole'' test case Eq.~\eqref{eq:tescase} with $J=10$, $g=1$, $H_\text{ext} = 0.05$, $T=10$. The input CF was generated in 10000 MC steps, with measurements every 
 13 steps. $L=64$ and $R_\text{cut}=16$ were used in both, the direct and the inverse MC algorithm. 
 $\Lambda = 0.10$.
 A random initial potential of small amplitude was ``preconditioned'' by local potential corrections (6996 MC steps) \cite{Almarza2003determination}.
 The correlation $C_r$ converged after 19 inverse MC iterations ($t^\dagger=50171$), yielding
 a deviation of $\vartheta^\dagger = 0.99$, cf. Tab.~\ref{tab:invMC_exp}.
 The reconstructed external field is $H_\text{ext} = 0.0477$.\\
 The double-logarithmic plot in the lower panel reveals that the $1/r^3$ behavior can be reconstructed.
 Negative values in $W_r$ occur at $r=1$ and $r\ge9.87$ (not shown in the log-log plot).
 }
 \label{fig:invMC_toymodel}
\end{figure*}

The empirical finding with this extended algorithm is, however, that the numerical stability and the speed of convergence may reduce notably
when $H_\text{ext}$ is allowed to vary (``float'').
A fixed value of $H_\text{ext}$ was thus used whenever possible.
For images with  specific magnetizations $m$ near 0 (white-pixel concentrations $c_\text{white}$ near 50\%), 
$|H_\text{ext}|$ is small and we
obtained suitable values for it by trial and error. This was the case for 4 SFM images.
In the opposite case (2 SFM images), where $m$ is not small, we could not obtain (guess) suitable values for
$H_\text{ext}$, because the system's behavior changes
heavily as the interaction potential is iterated. In these cases, $H_\text{ext}$ was 
kept floating in the IMC procedure. This is illustrated by the values given in Tab.~\ref{tab:invMC_exp}.

\section{Test case: Ising model with dipole-dipole  interaction}\label{app:test_case}

We tested the implementation of the IMC algorithm on a two-dimensional ferromagnetic Ising model with additional long-ranged repulsive $1/r^3$ interaction, as discussed before
in the context of ultrathin magnetic films \cite{PhysRevLett.75.950} which exhibit a clustered state at intermediate temperatures. 
The Hamiltonian of this system is given by
\begin{eqnarray}
 \hspace{-20pt}\mathcal{H}_\text{test}(\{\sigma\}) &\!=\!&
  -\frac{J}{2} \sum_{i,j\in\mathcal{G}|r_{ij} = 1} \sigma_i \sigma_j \nonumber \\ 
  &&+ g \sum_{i,j\in\mathcal{G}|i\neq j} \frac{\sigma_i \sigma_j}{\|\bm{r}_{ij}\|^3} 
  - H_\text{ext}\,M(\{\sigma\}). \label{eq:tescase}
\end{eqnarray}
The first term is a nearest-neighbor interaction (Ising model, ferromagnetic for $J>0$), the second term an out-of-plane dipole-dipole interaction
(repulsive for parallel spins), the third term is due to an external magnetic field.
This Hamiltonian corresponds to a pair potential
\begin{equation}
 W_{\bm{r}} = g\frac{1}{r^3} - \frac{J}{2}\left\{\begin{array}{l l}
                                     1 & \text{for } r=1
                                     \\
                                     0 & \text{else}
                                    \end{array}\right.
                                    ,
\end{equation}
with $r = \|\bm{r}\|$.

The results in Fig.~\ref{fig:invMC_toymodel} show that the IMC procedure is able to reconstruct the underlying pair-potential from CFs obtained from
a direct simulation. For larger distances and hence small  values, the reconstruction is limited by noise.
If the number of MC steps in the direct simulation of the input CF is increased, this noise level decreases due to the improved statistical precision of the input data.
Since (unlike for experimental data) the statistical error of the input CF can be precisely measured  in the direct MC simulation, $\vartheta\le1$ is reached by the IMC, cf. Eq.~\eqref{eq:conv_crit}.

\section{SFM Images}\label{app:SFM_images}
\begin{figure*}[p]
\centering
   \setlength\fboxsep{0pt}\setlength\fboxrule{1pt}
   \begin{tabular}{l r}
   \fbox{\includegraphics[width=.40\textwidth]{mrch15s2_1_ethanol_013.png}}
   &
   \fbox{\colorbox{gray}{\includegraphics[width=.40\textwidth]{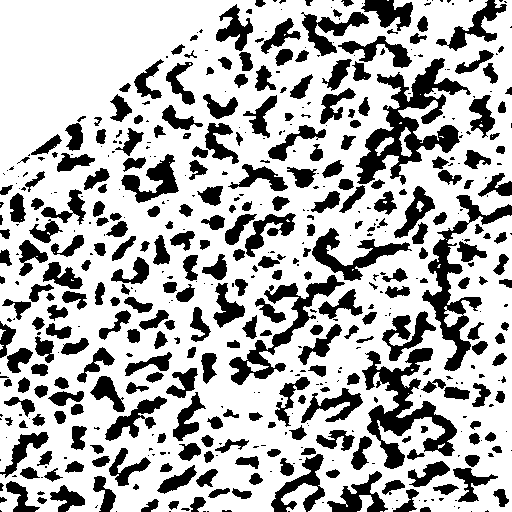}}}
   \\[3.6pt]
   \fbox{\includegraphics[width=.40\textwidth]{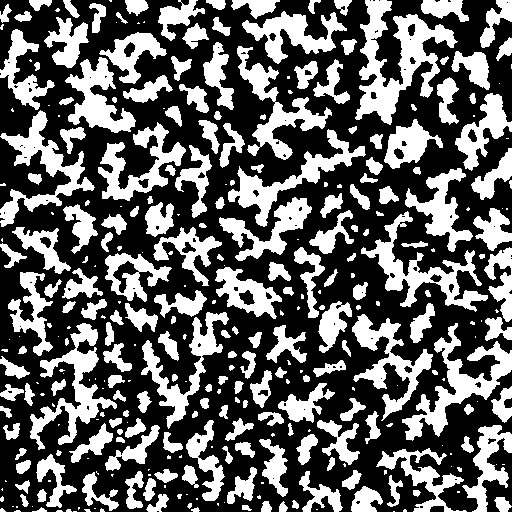}}  
   &
   \fbox{\colorbox{gray}{\includegraphics[width=.40\textwidth]{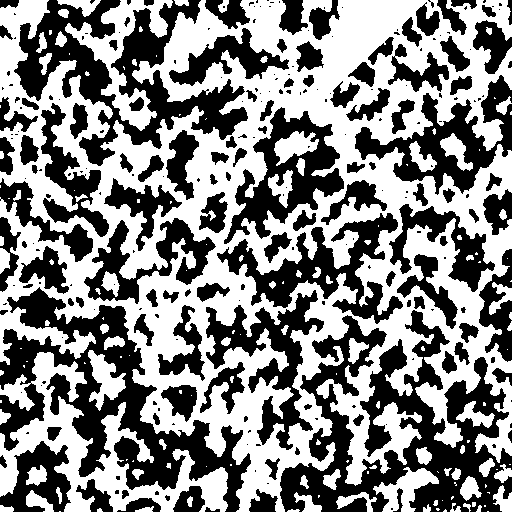}}}
   \\[3.6pt]
   \fbox{\includegraphics[width=.40\textwidth]{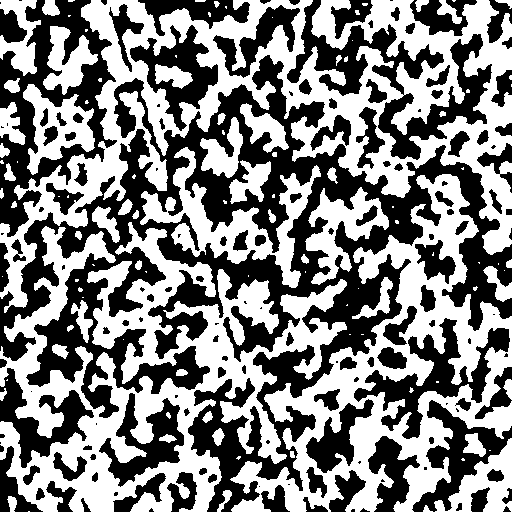}}
   &
   \fbox{\includegraphics[width=.40\textwidth]{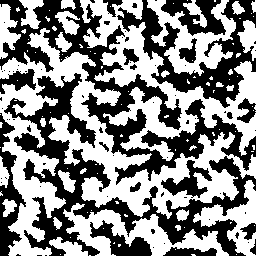}}
  \end{tabular}
  \caption[Analyzed experimental SFM images]{The six experimental SFM images analyzed in this work.
  Row-wise from top left to bottom right: G1\#1, G1\#2, G2\#1, G2\#2, G2\#3, G3\#1. Flattened, filtered and reduced to 1\, bit. Void pixels are shown in gray.
  G1\#1 is the image discussed in the main text.}
  \label{fig:SFM_images_SM}
\end{figure*}

\begin{table}[hb]
\caption[List of experimental SFM images]{
List of the six experimental SFM images analyzed in this work. $\sigma_\text{filter}$ is the width of the Gaussian convolution filter that was applied to the (flattened) raw image.
$c_\text{white}$ and $m$ are the white pixel percentage and the magnetization per volume for a spin system, respectively.
G1\#1 is the image discussed in the main text.
\\
* Image G3\#1 was not filtered, only flattened and reduced to 1\,bit. It was scaled down from 512\,pixel beforehand to simplify the inverse MC procedure.
}
\label{tab:SFM_images}
\centering
 \renewcommand{\arraystretch}{1.1}
 \footnotesize
 \begin{tabular}{|l| r| r|r| r| r| r|}\hline
  name 	& layers 	& width/px& $\sigma_\text{filter}$/px 	& width/nm 	& $c_\text{white}/\%$ 		& $m/\%$	\\\hline
  G1\#1	& 1 		& 512 		& 1.0					& 500 		&  48.6 	& $-2.9$	\\\hline
  G1\#2	& 1 		& 512 		& 1.0					& 533 		&  67.0 	& $+34.0$	\\\hline
  G2\#1	& 2 		& 512 		& 1.0					&  1000 	&  38.4 	& $-23.2$ 	\\\hline
  G2\#2	& 2 		& 512 		& 1.0					&  1000 	&  51.4 	& $+2.8$ 	\\\hline
  G2\#3	& 2 		& 512 		& 1.0					&  1000 	&  51.4 	& $+2.7$ 	\\\hline
  G3\#1	& 3 		& 256*		& 0.0					&  1000 	&  53.0 	& $+5.9$	\\\hline
 \end{tabular}
 \renewcommand{\arraystretch}{1.0}
\end{table}
The experimental SFM images that  were analyzed in this work are briefly described in Tab.~\ref{tab:SFM_images}.
The images themselves are shown in Fig.~\ref{fig:SFM_images_SM}.

Image G2\#3 shows a dark, slightly curled line from bottom middle to top left.
This is a depression in the graphene membrane of unclear origin. Possible causes:
\vspace{-6pt}
\begin{itemize} \itemsep2pt \parskip0pt \parsep0pt
 \item a defect in the mica crystal,
 \item a defect in the graphene sheet,
 \item a wrinkle (downward!) in the graphene possibly caused by compressive stress.
\end{itemize}
\vspace{-6pt}
In any case it noticeably affects the clustering patterns.

Gray pixels in the images indicate void areas, that were not included in the correlation functions.

\section{IMC Results}\label{app:IMC_results}
\begin{figure*}[hpt!]
 \centering
\begin{tabular}{p{.45\textwidth} p{.45\textwidth}}
\centering
\pdfpxdimen=\dimexpr 1in/96\relax
  \setlength\fboxsep{0pt}
  \setlength\fboxrule{1pt}
  \fbox{\includegraphics[clip=true, trim = 128px 128px 256px 256px, width=0.75\linewidth]{mrch15s2_1_ethanol_013.png}}
  &
  \centering
  \setlength\fboxsep{0pt}
  \setlength\fboxrule{1pt}
  \fbox{\includegraphics[width=0.75\linewidth]{./mrch15s2_1_ethanol_013_T1_00_H-0_001_conf.png}}
  \end{tabular}
  \begin{tabular}{p{.45\textwidth} p{.45\textwidth}}
    Detail of SFM image G1\#1, 128\,pixel  or 125\,nm wide.
    &
    Snapshot of MC simulation with the corresponding reconstructed potential and $H_\text{ext}=-0.001$ on a $128\times128$ lattice.
  \end{tabular}
  \caption{
  Comparison of experimental image G1\#1 and simulation with the reconstructed potential for this image.
  }
  \label{fig:invMC_example_conf}
\end{figure*}
\begin{figure*}[hpt!]
 \centering
  \includegraphics[clip = true, trim = 0 0 2mm 4mm,width = .88\linewidth]{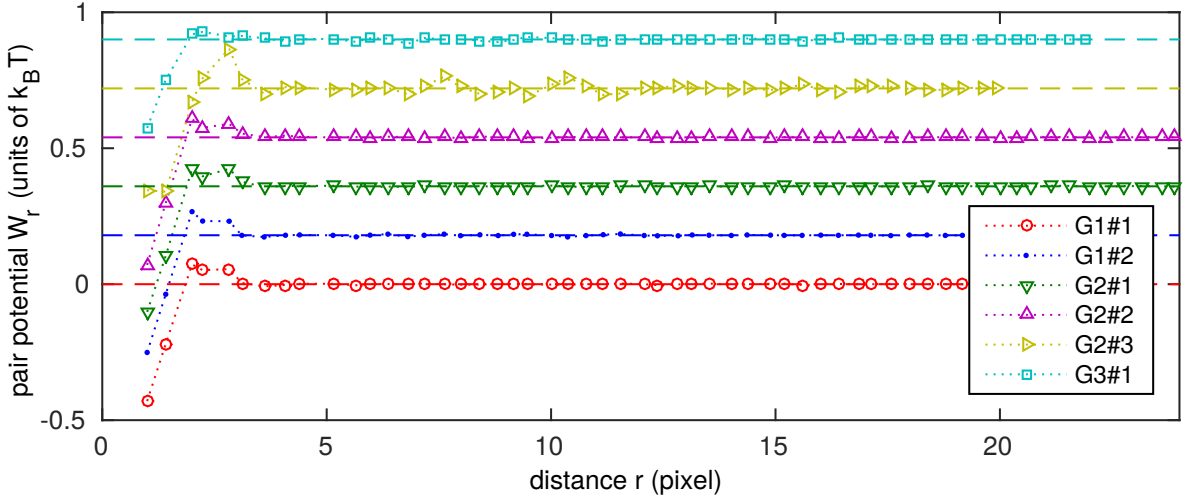}
  \caption{
  (Color online)
  Reconstructed potentials from SFM images: single- ($\mathrm{G}1\#n$), double- ($\mathrm{G}2\#n$)  and triple-layer ($\mathrm{G}3\#1$) graphene.
  Data is gradually shifted upward to improve readability. The corresponding zeros are indicated by dashed horizontal lines.
  Note that the $r$-axis is in pixels and that images have different resolutions, cf. Tab.~\ref{tab:SFM_images} and Fig.~\ref{fig:SFM_images_SM}.
  }
\label{fig:all_reconstr_pot_SM}
\end{figure*}
\begin{table*}[hpb!]
\centering
 \caption[Parameters for the inverse Monte Carlo algorithm with SFM images]{
 Parameters (middle part) and convergence data (right part) for the inverse Monte Carlo algorithm.
 $n_\text{iter}$ is the number of IMC iterations after which convergence occurred,
 $t^\dagger$ is the corresponding number of global MC updates. $\vartheta^\dagger$ is the relative
 correlation/magnetization deviation (cf. Eq.~\eqref{eq:conv_crit}) at the next measurement.
 $t_\text{CPU}$ is the  CPU time on one thread of an {Intel\textregistered} {Core\texttrademark} i7-3770 CPU @ 3.40GHz.\\
 The $H_\text{ext}$ values marked with an asterisk ($^*$) were fixed, the others are results from the algorithm.
 The specific magnetization $m$  (from Tab.~\ref{tab:SFM_images}) in the second column is shown once more to illustrate
 where $H_\text{ext}$ can be fixed ($|m|$ small) and where not ($|m|$ large).
 }
 \label{tab:invMC_exp}
 \small
 \begin{tabular}{|r| r || r |r | r| r|| r| r| r| r|}\hline
  image & $m/\%$	& $L$ 	& $R_\text{cut}$& $H_\text{ext}$& $\Lambda$ 	&$n_\text{iter}$& $t^\dagger/10^{5}$ 	&  $\vartheta^\dagger$  & $t_\text{CPU}$	\\\hline
  G1\#1 & $-2.9$	& 64 	& 22		& $-0.001^*$	& 0.50		& 14 		& $7.9$			& 0.98			& 12\,h 		\\\hline
  G1\#2 & $+34.0$	& 80 	& 20		& $+0.0056$ 	& 0.10		& 14 		& $3.7$			& 1.96			& 8\,h  		\\\hline
  G2\#1 & $-23.2$ 	& 80  	& 24		& $-0.0035$	& 0.15		& 23 		& $19.5$ 		& 1.99			& 56\,h 		\\\hline
  G2\#2 & $+2.8$ 	& 80  	& 24 		& $+0.001^*$	& 0.15		& 14 		& $6.3$			& 2.00 			& 18\,h 		\\\hline
  G2\#3 & $+2.7$ 	& 80  	& 20 		& $+0.000^*$	& 0.10  	& 27 		& $11.8$		& 1.05  		& 24\,h 		\\\hline
  G3\#1 & $+5.9$	& 89 	& 22		& $+0.001^*$	& 0.50		& 30 		& $5.7$			& 2.85  		& 18\,h 		\\\hline
 \end{tabular}
\end{table*}

In Fig.~\ref{fig:invMC_example_conf}, we compare the original SFM image with a spin configuration from a MC simulation with the reconstructed potential.
Even though any quantitative judgment should involve the correlation function, this depicts the similarity between experimental and reconstructed 
system.

The reconstructed potentials for all six SFM images are shown in Fig.~\ref{fig:all_reconstr_pot_SM}. The parameters used for the IMC procedure are 
listed in Tab.~\ref{tab:invMC_exp}, along with the computing times and the deviations from the experimental data that was reached.
Remarkable features that all  potentials have in common:
short-ranged attractive, then somewhat longer-ranged 
repulsive and  apparently fluctuating around 0 for larger distances.
The potentials from G1\#1--G2\#2 show a very similar pattern:
the first two values are negative, where $W_1\approx 1.8\,W_{\!\!\sqrt{2}}$, then four positive values in a characteristic 
zigzag pattern with the most repulsive value at the third 
shortest distance $r=2$ then apparently unsystematic fluctuations around $W=0$.
Deviations occur for G2\#3, which has a dark line across the image (cf. Fig.~\ref{fig:SFM_images_SM}) and the triple-layer image
G3\#1, which has the largest structures of all images.

At this point, we  would once again like to point out the geometrical meaning of
distances on the lattice.
Fig.~\ref{fig:invMC_lattice_pot_SM} shows the reconstructed potential from G1\#1 on the lattice for short distances.
One can clearly see a square-shaped potential barrier. This certainly has a geometrical 
meaning connected with the lattice geometry and the implementation of certain concepts on a lattice, e.\,g., an elastic line-tension
that we expect to be present due to the elastic graphene deformation at the cluster boundaries.

Furthermore, the results are effective potentials. One should thus \emph{not interpret} such a result as: ``The microscopic, intermolecular potential is attractive at a distance of $\sqrt{2}$\,pixel = 1.38\,nm 
and repulsive at a distance of ${2}$\,pixel = 1.95\,nm.''

\subsection*{Possible Error Sources}
There remains some uncertainty about the scope of application of the reconstructed potentials.
First of all, the SFM images are prone to noise and thermal drift. Hence we
corrected for the drift, applied a Gaussian filter and reduced the color depth.
This manipulation to the raw image could introduce an error. However, we analyzed the 
influence on the correlation function \cite{severin2015nanophase} with the finding that it is quite stable to 
the image manipulation.

Secondly, the statistical basis obtained from one image ($512\times512$\,pixel) could be too poor
to extract even the main characteristics of the interaction potential. In addition, our estimate of
the input correlation function's error is very coarse.

\section{MC Study of the Resulting System}\label{app:MC_study}
\begin{figure*}[t]
\centering
\begin{tabular}{p{.45\textwidth} p{.45\textwidth}}
  {\includegraphics[clip = true, trim = .2pt 6pt 6pt 17pt,scale=.96]{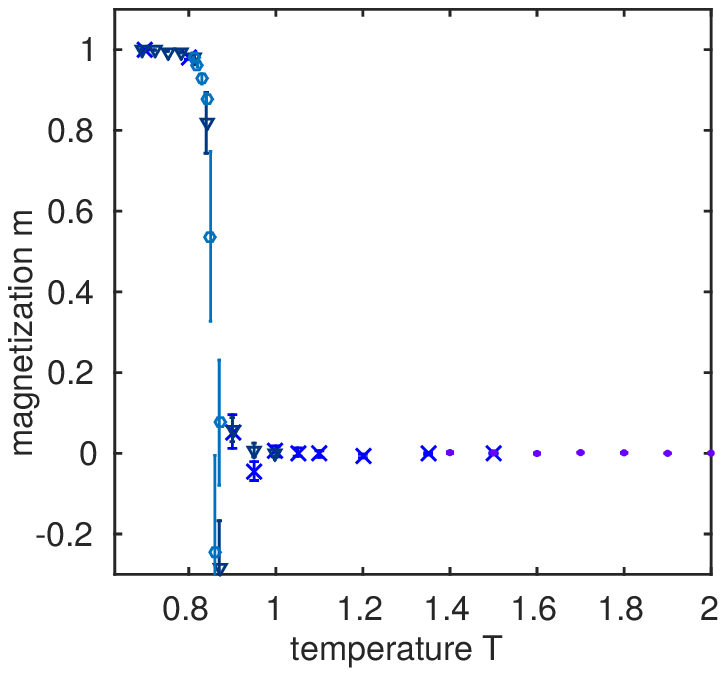}}&
  {\includegraphics[clip = true, trim = .2pt 4.5pt 6pt 15pt,scale=.96]{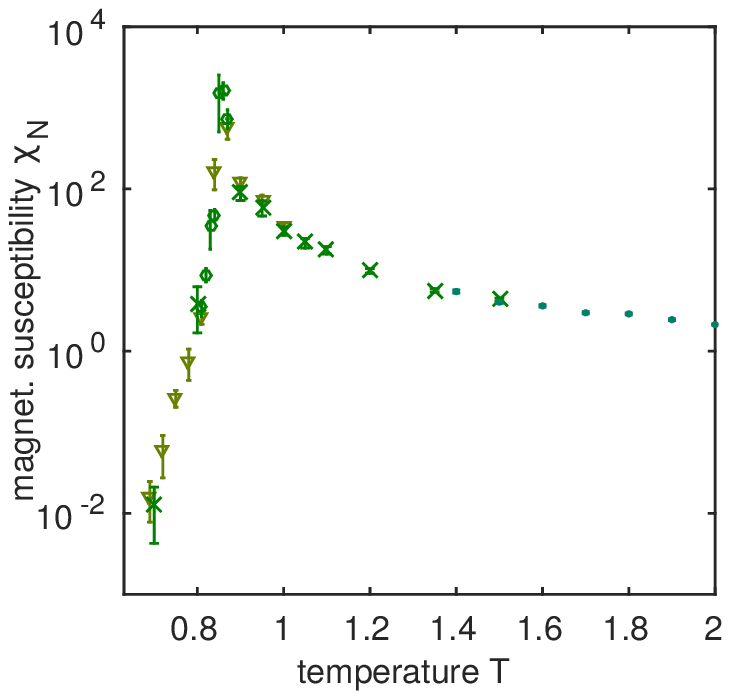}}
\end{tabular}  
 \caption[MC simulations with reconstructed potential, variable $T$: Observables]{
  (Color online)
  MC simulations with reconstructed potential (Fig.~\ref{fig:invMC_example_pot}): Observables $m$ (left panel) and $\chi_N$ (right panel) in dependence on $T$
  with fixed $H_\text{ext}=0.0$. 
  Measured with decreasing $T$ in four independent series (different markers)
  on a $64\times 64$ lattice, $10^4$ MC updates per measurement
  (of which 10\% for thermalization; $T\in[0.81, 0.87]$ was resolved with $10^5$ MC updates per measurement).  
  }
  \label{fig:reconst_pot_obs_T_SM}
\end{figure*}

In direct MC simulations using the reconstructed potential, we examined the three observables specific heat capacity, (specific) magnetization, and
magnetic susceptibility
\begin{eqnarray}
  &c_N& = \frac{1}{N} \left(\frac{\partial  E }{\partial T}\right)_N
 = \frac{\langle \mathcal{H}^2\rangle-\langle \mathcal{H}\rangle^2}{N\,T^2},\\
 &m& = \frac{M}{N}= \frac{1}{N}\sum_{i\in\mathcal{G}}\sigma_i\\
 &\chi_N& = \frac{1}{N} \left(\frac{\partial  M }{\partial H_\text{ext}}\right)_{N,T}
 =\frac{\langle{M}^2\rangle-\langle {M}\rangle^2}{N\,T}.
\end{eqnarray}
We used the potential  G1\#1 to obtain all the data shown here. However, simulations with G2\#1 and G3\#1 yielded similar results.

\subsection[Variable External Field]{Variable Field $H_\text{ext}$}
We fixed $T=1$ and ramped $H_\text{ext}$ up and down.
 No hysteresis could be detected in the range $H_\text{ext}\in[0.00, 0.15]$.
The magnetization $m$  sensitively increases with $H_\text{ext}$ and shows saturation already at values below $H_\text{ext}=0.1$.
Correspondingly, the susceptibility $\chi_N$ is high.
The external field depends on the chemical potential
in the binary mixture picture. 
Hence a
difference between $V_{AA}$ and $V_{BB}$ (in the sense of Eq.~\eqref{eq:binmixtHext}) by less than $0.2\,k_\text{B}T$
leads to a system dominated by one of the components.
Back in the spin picture, we are obviously dealing with a paramagnetic phase here (i.\,e. near  $T=1$).

\subsection{Variable Temperature $T$}
\begin{figure*}[ht]
  \centering
  \includegraphics[clip=true,trim=28mm 16mm 24mm 10mm,  width=.9\textwidth]{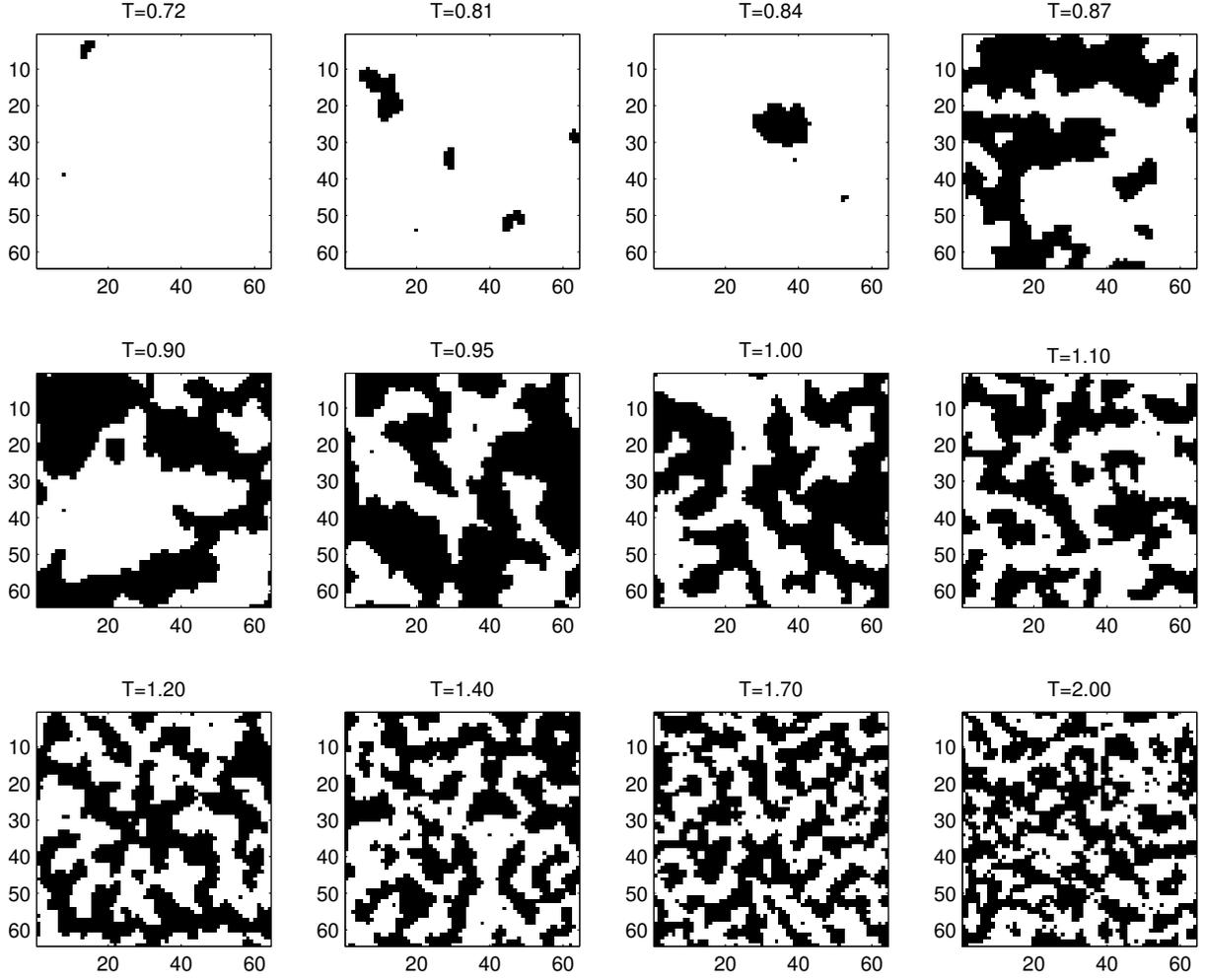}
  \caption[MC simulations with reconstructed potential: Snapshots of spin states]{
    Snapshots of spin states from  MC simulations 
    with the reconstructed potential for image G1\#1. Selected temperatures $T$; $64\times 64$ lattice,
    cf. Fig.~\ref{fig:reconst_pot_obs_T}.
  }
  \label{fig:reconst_pot_conf_SM}
\end{figure*}

The temperature dependence of $c_N$ for $H_\text{ext}=0$ is shown in Fig.~\ref{fig:reconst_pot_obs_T} and we add here the dependence of $m$ and $\chi_N$ in Fig.~\ref{fig:reconst_pot_obs_T_SM}.
Snapshots of spin configurations at different temperatures are shown in Fig.~\ref{fig:reconst_pot_conf_SM}. One can clearly see the phase transition
happening between $T=0.84$ and $T=0.87$, where the black-white symmetry is broken which means spontaneous magnetization.

\subsection{Results for Thicker Graphenes}
\begin{figure*}[ht]
 \centering
 {\includegraphics[clip=true, trim=83pt 0 85pt 8pt, width=\textwidth]{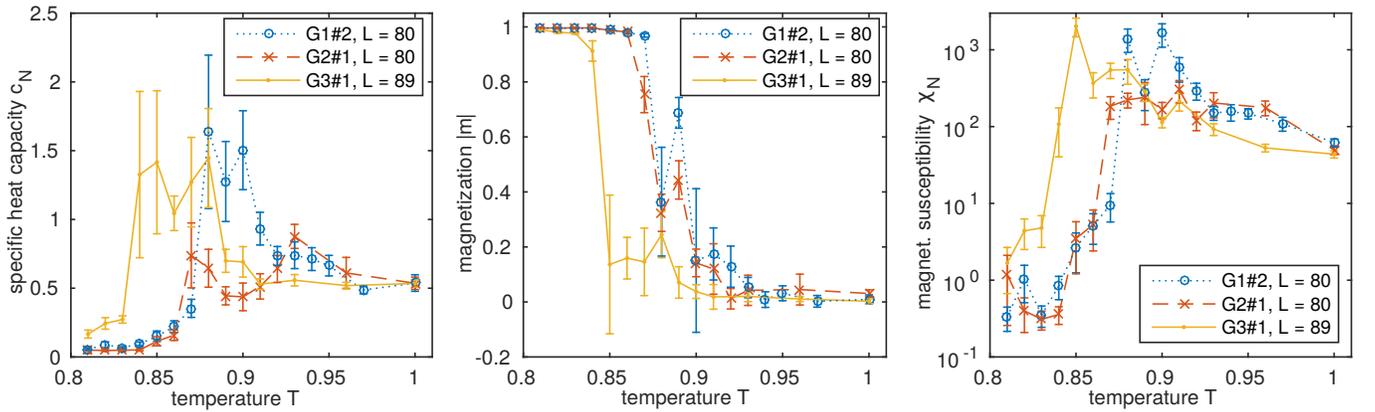}}
 \caption{
  (Color online)
  MC simulations for the potentials (cf. Fig.~\ref{fig:all_reconstr_pot_SM}) reconstructed from images G1\#2, G2\#1 and G3\#1 (single-, double- and triple-layer, respectively):
  Observables $c_N$ (left panel), $m$ (middle) and $\chi_N$ (right panel) in dependence on $T$
  with fixed $H_\text{ext}=0.0$. 
  Measured with decreasing $T$. The absolute value of the averaged magnetization $|\langle m\rangle|$ is plotted instead of the magnetization $\langle m\rangle$, because its sign is arbitrary
  in the absence of an external field.\\
  The critical temperatures are $T_c=0.88$ for G1\#2, $T_c\in[0.86, 0.88]$ for G2\#1 and $T_c=0.85$ for G3\#1, respectively.
 }
 \label{fig:123layers_T_c}
\end{figure*}

Apart from the potential for image G1\#1 (single-layer graphene; $m\approx0$, i.\,e., black/white symmetric) discussed in the main text,
we also performed MC simulations with the potentials for images G1\#2 (single-layer, $m>0$), G2\#1 (double-layer, $m<0$) and G3\#1 (triple-layer, $m\approx0$). In all cases the 
external field was set to $H_\text{ext}=0$ and the temperature was decreased from $T=1$ to observe the critical point.
Results for the thermodynamic observables are shown in  Fig.~\ref{fig:123layers_T_c}. The   critical temperatures ($T_c=0.88$ for G1\#2, $T_c\in[0.86, 0.88]$ for G2\#1 and $T_c=0.85$ for G3\#1)
show no systematic dependence on the number of graphene layers.

\bibliography{bibliography_final}

\end{document}